\documentstyle[12pt,psfig]{article}
\makeatletter
\def\@cite#1#2{{[{#1}]\if@tempswa\typeout
{IJCGA warning: optional citation argument
ignored: `#2'} \fi}}
\newcount\@tempcntc
\def\@citex[#1]#2{\if@filesw\immediate\write\@auxout{\string\citation{#2}}\fi
  \@tempcnta\z@\@tempcntb\m@ne\def\@citea{}\@cite{\@for\@citeb:=#2\do
    {\@ifundefined
       {b@\@citeb}{\@citeo\@tempcntb\m@ne\@citea\def\@citea{,}{\bf ?}\@warning
       {Citation `\@citeb' on page \thepage \space undefined}}%
    {\setbox\z@\hbox{\global\@tempcntc0\csname b@\@citeb\endcsname\relax}%
     \ifnum\@tempcntc=\z@ \@citeo\@tempcntb\m@ne
       \@citea\def\@citea{,}\hbox{\csname b@\@citeb\endcsname}%
     \else
      \advance\@tempcntb\@ne
      \ifnum\@tempcntb=\@tempcntc
      \else\advance\@tempcntb\m@ne\@citeo
      \@tempcnta\@tempcntc\@tempcntb\@tempcntc\fi\fi}}\@citeo}{#1}}
\def\@citeo{\ifnum\@tempcnta>\@tempcntb\else\@citea\def\@citea{,}%
  \ifnum\@tempcnta=\@tempcntb\the\@tempcnta\else
   {\advance\@tempcnta\@ne\ifnum\@tempcnta=\@tempcntb \else
\def\@citea{--}\fi
    \advance\@tempcnta\m@ne\the\@tempcnta\@citea\the\@tempcntb}\fi\fi}
\makeatother

\def\bma#1{\mbox{\boldmath{$#1$}}}

\newcommand{\gsim}{\lower.7ex\hbox{$\;\stackrel{\textstyle>}{\sim}\;$}}
\newcommand{\lsim}{\lower.7ex\hbox{$\;\stackrel{\textstyle<}{\sim}\;$}}
\newcommand{\be}{\begin{equation}}
\newcommand{\ee}{\end{equation}}
\newcommand{\bea}{\begin{eqnarray}}
\newcommand{\eea}{\end{eqnarray}}
\def\simlt{\stackrel{<}{{}_\sim}}

\def\baselinestretch{1}
\begin{document}
%%%%%%%%%%%%%%%%%%%%%%%%%%% subequations.sty %%%%%%%%%%%%%%%%%%%%%%%%
\catcode`@=11
\newtoks\@stequation
\def\subequations{\refstepcounter{equation}%
\edef\@savedequation{\the\c@equation}%
  \@stequation=\expandafter{\theequation}%   %only want \theequation
  \edef\@savedtheequation{\the\@stequation}% % expanded once
  \edef\oldtheequation{\theequation}%
  \setcounter{equation}{0}%
  \def\theequation{\oldtheequation\alph{equation}}}
\def\endsubequations{\setcounter{equation}{\@savedequation}%
  \@stequation=\expandafter{\@savedtheequation}%
  \edef\theequation{\the\@stequation}\global\@ignoretrue

\noindent}
\catcode`@=12
%%%%%%%%%%%%%%%%%%%%%%%%%%%%%%%%%%%%%%%%%%%%%%%%%%%%%%%%%%%%%%%%%%%%%
\begin{titlepage}

\title{{\bf  Effective operators and vacuum instability
as heralds of new physics}}
\vskip2in
\author{
{\bf C.P. Burgess$^{1}$\footnote{\baselineskip=16pt E-mail: {\tt
cliff@physics.mcgill.ca}}}, {\bf V. Di
Clemente$^{2}$\footnote{\baselineskip=16pt E-mail: {\tt
vicente@hep.phys.soton.ac.uk}}} and {\bf J.R.
Espinosa$^{3,4}$\footnote{\baselineskip=16pt E-mail: {\tt
espinosa@makoki.iem.csic.es}}}
\hspace{3cm}\\
%\vskip.35in
 $^{1}$~{\small McGill Univ., 3600 University St., Montr\'eal, Qu\'ebec,
Canada, H3A 2T8}
\hspace{0.3cm}\\
 $^{2}$~{\small Univ. of Southampton,
Southampton, SO17 1BJ, U.K.}
\hspace{0.3cm}\\
 $^{3}$~{\small I.M.A.F.F. (CSIC), Serrano 113 bis, 28006 Madrid, Spain}
\hspace{0.3cm}\\
 $^{4}$~{\small I.F.T. C-XVI, U.A.M., 28049 Madrid, Spain}
}
\date{}
\maketitle
\def\baselinestretch{1.15}
\begin{abstract}
\noindent For a light enough Higgs boson, the effective potential
of the Standard Model develops a dangerous instability at some
high energy scale, $\Lambda$, signalling the need for new physics
below that scale. On the other hand, a typical low-energy remnant
of new physics at some heavy scale, $M$, is the presence of
effective non-renormalizable operators (NROs), suppressed by
powers of $1/M$. It has been claimed that such operators may
modify the behaviour of the effective potential, in such a way as
to significantly lower the instability scale. We critically
reanalyze the interplay between non-renormalizable operators and
vacuum instabilities and find that, contrary to these claims, the
effect of NROs on instability bounds is generically small whenever
it can be reliably computed.
\end{abstract}

\thispagestyle{empty}
\vspace{4cm}
\leftline{December 2001}
\leftline{}

\vskip-22cm
\rightline{}
\rightline{McGill-01/27}
\rightline{SHEP 02-01}
\rightline{IFT-UAM/CSIC-01-41}
\rightline{hep-ph/0201160}
\vskip3in

\end{titlepage}
%%%%%%%%%%%%%%%%%%%%%%%%%%%%%%%%%%%%%%%%%%%%%%%%%%%%%%%%%%%%%%%%%%%
\setcounter{footnote}{0} \setcounter{page}{1}
\newpage
\baselineskip=20pt

\noindent

It is well known that the Higgs effective potential in the Standard
Model (SM) develops an instability if the Higgs mass is below some
critical value, termed the vacuum stability bound
\cite{bounds,stab,altarelli,Boya}. This
potential instability appears due to the fact that the Higgs quartic
coupling, $\lambda$, which is positive at the electroweak scale (where it
determines the Higgs mass), can run towards negative values in the
ultraviolet (due to radiative corrections from top-quark loops). If this
happens at some high energy scale $\Lambda$, for values of the Higgs field
$H\sim \Lambda$, the potential is dominated by the negative $\lambda |H|^4$
term and it is either unbounded from below or develops a very deep minimum
(deeper than the electroweak vacuum) beyond $\Lambda$.

For a given cut-off scale $\Lambda$ one can compute the critical value of the Higgs mass,
$M_h^*(\Lambda)$, below which the potential will develop an
instability\footnote{Weaker (metastability) bounds on $M_h$ result if
one accepts
a non-standard minimum deeper than the electroweak vacuum provided that
the lifetime
(for quantum mechanical or thermally activated decay) of the latter is longer
than the age of the Universe \cite{meta,strumia}.} at,
or below, $\Lambda$. As an example, for a top-quark mass $M_t=175$ GeV,
the stability bound is $M_h^*\simeq 50$ GeV for $\Lambda=1$~TeV and
$M_h^*\simeq 130$ GeV for $\Lambda=10^{19}$ GeV \cite{stab}.
Alternatively, for $M_t=175$ GeV and $M_h=115.6$ GeV (as suggested
by LEP2 \cite{LEP}), the Higgs potential of the Standard Model develops an
instability at the scale $\Lambda^*\simeq 100$~TeV \cite{stab}.

The significance of these analyses is that such an instability
provides circumstantial evidence for the presence of new physics
at energies which cannot be made arbitrarily large compared to the
electroweak scale, $v = 246$ GeV. This evidence relies on the
modern picture of the SM as the low-energy approximation to a more
complete theory, with the full theory deviating in its predictions
from those of the SM only by powers of $E/M$, where $E$ is the
energy of the observable of interest, and where $M$ is the mass
scale of the new physics. Viewing the SM prediction of vacuum
instability as a failure to reproduce the vacuum properties of
this more complete theory for energies $E \sim \Lambda^*$, we must
conclude that the parameter controlling the difference between
these two theories, $\Lambda^*/M$, cannot be too small.

Of course $\Lambda^*$ as defined by the stability analysis only
gives an indication of where the scale, $M$, of new physics must
lie, and does not provide a strict upper bound. It is quite
possible that the scale $M$ -- defined, say, as the mass of the
lightest hitherto-undiscovered particle -- is a bit larger than
$\Lambda^*$ \cite{HS,vicente}. For instance, in the examples
explored by Ref.~\cite{vicente} (using a toy model with parameters
that mimic those of the MSSM) the mass-scale\footnote{In this
kind of analysis it is
necessary to deal with a multi-scale problem due to the presence
of several mass scales ({\it e.g.} the Standard Model scale and
the new physics scale) in the effective
potential~\cite{multiscale}.} of the perturbative
new physics can be as large as $M = 4\Lambda^*$.

It is well known that integrating out new physics having mass $M$
generates a host of non-renormalizable operators (NROs) in the
effective theory, which are suppressed by powers of $1/M$ and
which express in detail how the full theory differs from the SM at
low energies, $E\ll M$. It has been remarked by several authors
that such operators also contribute to the Higgs potential, $V$,
in a way which might affect the vacuum stability bounds. For
instance the appearance of a NRO of the form
\be \delta V = -{\alpha\over 3 M^2}|H|^6\ , \label{NRO} \ee
changes the behaviour of $V$ by an amount which is small for small
$H$, but which might nonetheless successfully compete with equally
small perturbative SM effects. By including such a term in the
Higgs potential, refs.~\cite{datta,wudka} in this way find the
vacuum instability to arise for scales $\Lambda^*_\alpha$ which
can be a few orders of magnitude smaller than in the absence of terms
like eq.~(\ref{NRO}).

It is our purpose in this letter to critically re-examine these
arguments. We argue that large changes to $\Lambda^*$ typically
indicate a breakdown of the approximations being used, rather than
providing a solid indication that new physics must exist at
comparatively low energies. For simplicity we limit our analysis
to a single non-renormalizable operator of the form (\ref{NRO}),
and take $\alpha\geq 0$ (which is the case of interest because, as
we will see, it tends to lower the instability scale). Although
one generally expects a host of different NROs to arise when
integrating out heavy physics, our main conclusions are not
substantially affected by the presence of these other effective
operators.

We present our arguments in the following way. In section~1 we
reproduce previous studies which find sizable changes to
$\Lambda^*$ due to this interplay between stability bounds and
non-renormalizable operators. In section~2, we explain,
with the help of a toy
model, the dangers of using an effective theory (valid below some
cut-off) to study properties of the effective potential at values
of the field close to the cut-off.

In section~3 we present our alternative, 'bottom-up', approach to
this problem: assuming that we know $M_h$ and have indications of
the existence of a NRO like (\ref{NRO}), we first examine how the value
of $M$ in (\ref{NRO}) can be estimated. We then show how
the calculation of the instability scale $\Lambda^*$ changes due
to the NRO of eq.~(\ref{NRO}). We consider in turn the three
mutually-exclusive and exhaustive cases for the relative sizes of
the new physics scales, $M$ and $\Lambda^*$: {\it (a)}
$\Lambda^*\gg M$, {\it (b)} $\Lambda^*\simeq M$ and {\it (c)}
$\Lambda^*\ll M$. [Here $\Lambda^*$ is the instability scale in
the pure SM, without NROs, and $M$ is the new physics scale
appearing in (\ref{NRO})]. We show that only case {\it (c)} can be
studied reliably in an effective theory approach, and we conclude
that the change in $\Lambda^*$ due to the NRO (\ref{NRO}) is in
general small when its effects are reliably calculable. We discuss
separately the particular choices of parameters -- $M_h\sim
120-130$ GeV (the precise value depends on $M_t$ and the strong
coupling constant $\alpha_s$) -- that require some qualifications,
but which do not change our conclusion.

{\bf 1. Previous Analyses:} References \cite{datta,wudka} study in
detail the possible influence of non-renormalizable operators (in
the Higgs potential) on vacuum stability bounds. Although they
proceed with different levels of sophistication ({\it e.g.}
ref.~\cite{wudka} includes the influence of NROs on the
renormalization group evolution of the parameters, an effect that
was not considered in ref.~\cite{datta}) both find that
non-renormalizable terms, like the one presented in
eq.~(\ref{NRO}), do have a significant impact on stability bounds.

There are two (equivalent) ways of stating their results:

\begin{itemize}
\item {\it A)} For a given value of the cut-off scale $\Lambda$, the
corresponding value of the stability bound on the Higgs mass,
$M_h^*(\Lambda)$, can be relaxed ({\it i.e.} the bound increases) quite
significantly due to the presence of
non-renormalizable operators like (\ref{NRO});

\item {\it B)} For a fixed value of the Higgs mass, the scale $\Lambda^*$
at which the Higgs potential develops an instability can be lowered
dramatically
by the presence of non-renormalizable operators like (\ref{NRO}).
\end{itemize}

These results are obtained as follows. Let us assume for
simplicity that the only NRO present is of the form given by
equation~(\ref{NRO}), with $\alpha>0$ (as already mentioned, we
focus on this sign for $\alpha$ because we are more interested in
effects that lower the scale of new physics). For a fixed
$\Lambda$, the stability bound in the pure SM, $M_h^*(\Lambda)
\equiv M_h^*(\Lambda, \alpha=0)$, is obtained 
\cite{bounds,stab,altarelli,Boya} by setting
$\lambda(Q=\Lambda)\simeq 0$ (where $Q$ is the renormalization
scale) as a boundary condition. More precisely we require
\be \left.\hat\lambda(Q)\equiv
\left[\lambda(Q)+ \delta\lambda\right]
\right|_{Q=\Lambda} = 0\ , \label{bound}
\ee
where $\delta\lambda$ is a (non-logarithmic) correction of
one-loop order. In our definion, $\hat\lambda(Q)$ is simply the
quartic Higgs coupling in the one-loop potential evaluated at
$Q=|H|$ and it tracks the zeroes of $V$ for large $|H|$. This
quantity differs slightly from a similar one-loop corrected
$\tilde{\lambda}(Q)$ used in Ref.~\cite{stab} that tracks instead the
extrema of $V$. Condition (\ref{bound}) ensures
$V(Q=\Lambda)\simeq \hat\lambda(\Lambda)|H|^4/2\simeq 0$ near
$|H|\sim\Lambda$.

For $\alpha\neq 0$ and positive, however, one has to deal also
with the $|H|^6$-term, which may be competitive with the
$|H|^4$-term in spite of the $M^{-2}$ suppression of the former.
If $M\simeq \Lambda$ (more about this later) we are led to the
approximate condition (for borderline stability)
\be {1\over 2}\hat\lambda_\alpha(Q)\equiv
\left[ {1\over
2}\hat\lambda(Q)-{1\over 3}\alpha(Q)\left.{|H|^2\over
\Lambda^2}\right]\right|_{Q=\Lambda}\simeq 0\ , \ee
for $|H|\simeq \Lambda$. That is, the boundary condition for
$\lambda$ will be
\be \lambda(\Lambda)\simeq
-\delta\lambda+{2\over 3} \alpha(\Lambda) = 0\ . \label{bounda}
\ee
Comparing (\ref{bounda}) to (\ref{bound}), and noting that a
larger value of $\lambda$ at the scale $\Lambda$ translates into a
larger value at the electroweak scale, we arrive at $M_h^*(\Lambda,\alpha)\geq
M_h^*(\Lambda,\alpha=0)$. This conforms to statement {\it (A)}
above. For $\Lambda\simlt 50$ TeV, the increase in
$M_h^*(\Lambda,\alpha)$ for moderate values of $\alpha$ is claimed
\cite{wudka} to be $\sim 40-60$ GeV, which is a rather large
effect.

Alternatively, for a fixed value of $M_h$, the instability scale
in the SM, $\Lambda^*(M_h)$, is determined by condition (\ref{bound}). For a
non-zero $\alpha>0$, however, the potential seems to run into an
instability as soon as $\lambda(Q)$ is so small that $\lambda
|H|^4$ cannot compete any more with the negative term $-\alpha
|H|^6$. So, the new instability scale must satisfy
$\Lambda^*_\alpha \equiv \Lambda^*(M_h,\alpha)< \Lambda^*(M_h,
\alpha=0)$ as stated in {\it (B)} above. For $M_h\simeq 115.6$ GeV
(as suggested by LEP \cite{LEP}) the instability scale in the pure SM,
$\Lambda^*\simeq 100$ TeV,  would be lowered by this
argument, according to \cite{wudka}, to $\Lambda^*_\alpha\simeq
1-20$ TeV depending on $\alpha$. This is again a rather dramatic
downward shift.

{\bf 2. The Expansion of ${\bma V}$ in Powers of
${\bma H}{\bma /}{\bma M}$:} The large size
of the corrections obtained by the previous arguments is troubling
from the point of view of the effective theory, since the validity
of the low-energy expansion itself usually ensures that the effects
of NROs are small corrections to the predictions of the
renormalizable low-energy theory.\footnote{Of course an important
exception to this statement arises for observables for which the
predictions of the renormalizable theory are themselves small,
such as when these are suppressed by a conservation law or
selection rule of the renormalizable theory.} If NROs
substantially can change the predictions of the SM in this
instance, we must ask why the same is not true in other situations
for which the SM seems to work well. In other instances where 
similarly large results from effective NROs were found, the conclusions
turned out to be artifacts of applying the effective theory outside
its domain of validity \cite{BL}.

For an indication of what is going on, recall the assumption made
in the derivation that $H \simeq \Lambda \simeq M$, since one
might under these circumstances question the expansion of the
scalar potential in powers of $H/M$. Indeed, the dangers of making
this kind of expansion in a stability analysis may be illustrated
by the following toy model. Consider two real scalar fields,
$\phi$ and $\Phi$, with potential
\be V(\phi,\Phi)=-{1\over 2}m^2\phi^2+{1\over
8}\lambda\phi^4+{1\over 2}M^2\Phi^2 +\xi
\phi^3\Phi+\kappa\phi^2\Phi^2\ , \label{fullpot} \ee
and take $M^2\gg m^2 > 0$. The light field, $\phi$, acquires a
vacuum expectation value (VEV), $\langle\phi\rangle\sim -m^2/
\lambda$ and this also triggers a non-zero, but small, VEV for
$\Phi$, but these are complications inessential for our purpose.

To ensure that the the potential (\ref{fullpot}) is not unbounded
from below we must choose our parameters in such a way as to
ensure
\be \frac{1}{8}\lambda + \xi y + \kappa y^2 \geq 0 \label{UFB2}
\ee
where $y\equiv \Phi/\phi \in (-\infty, \infty)$. This condition is
automatically satisfied provided
\be \kappa>0\ ,\;\;\;\; {\mathrm
and}\;\;\;\; \lambda \geq 2 \frac{\xi^2}{\kappa}\ , \label{UFB3}
\ee
which is a condition we assume to hold in what follows.

For energies small compared to $M$ we can integrate out the heavy
field $\Phi$ to obtain the low-energy potential, which at tree
level becomes
\be V(\phi)=-{1\over 2}m^2\phi^2+{1\over 8}\lambda\phi^4-{1\over
2}\xi^2{\phi^6\over M^2+2\kappa\phi^2} \ . \label{pot} \ee
If this is expanded in powers of $\phi^2/M^2$, then the dominant
term is precisely of the form (\ref{NRO}), with $\alpha = 3\,
\xi^2 /2>0$. If we were to truncate the potential (\ref{pot}) at
$O(\phi^6/M^2)$ and ignore higher powers of $\phi$, we would
conclude that the potential develops an instability when
\be {\phi^2\over M^2}\simeq {\lambda\over 4\xi^2}\ ,
\label{phiMtoy} \ee
due to the negative $\phi^6$-contribution. Using (\ref{UFB3}), it
would be tempting to justify the truncation of the potential on
{\it a posteori} grounds, since (\ref{phiMtoy}) implies
$(\phi/M)^2 \geq 1/(2\kappa)$, which can be smaller than 1 for
$\kappa\geq 0.5$.

The instability found in this way is specious, however, because we
know that it is an artifact of the truncation of the potential to
order $\phi^6/M^2$. This is a bad approximation to the potential,
(\ref{pot}), which does not share this instability by virtue of
the second of conditions (\ref{UFB3}), which guarantees the
stability of the underlying potential, (\ref{fullpot}). For values
of $\phi$ close to the scale $M$, it is not correct to expand the
potential within the effective theory, even if the effective
theory itself {\it is} perfectly well justified (such as by having
$m \ll M$).

How can it be that the effective theory can be justified as an
expansion in powers of $1/M$ and yet the potential cannot also be
so expanded? This is not so odd as it might seem. The validity of
the effective theory rests on the absence of large energy
densities, since these might allow the production of the heavy
particles which have been integrated out. In our example this
justifies the suppression of terms involving powers of $m/M$, and
of terms involving higher derivatives divided by powers of $M$.
The validity of the expansion of $V(\phi)$ in powers of $\phi/M$
hinges on whether or not fields for which $\phi/M \simeq 1$ can
produce an energy density which is small enough to remain within
the low-energy regime.

That is, if the scalar potential is sufficiently shallow as a
function of $\phi$, it can be that $\phi \simeq M$ and yet
$V(\phi)$ remains small. If so, it is illegitimate to expand
$V(\phi)$ in powers of $\phi/M$ even within the low-energy theory.
This situation frequently arises within supergravity theories, for
which the scalar potentials often have flat directions. In such
cases one keeps in the effective theory the terms with the fewest
derivatives (such as the Einstein-Hilbert action) but also finds
nonpolynomial scalar potentials.

The Higgs potential of the SM does not have such a flat direction,
since the quartic interaction generically ensures that fields
$H\simeq M$ produce energy densities of order $V(H) \simeq \lambda
M^4$. This is why it is generally a good approximation to expand
the SM potential in powers of $H/M$, keeping only the
renormalizable terms. The exception to these statement arises
precisely in the instance of marginal stability, when the quartic
coupling is close to vanishing: the case of interest of the
previous analysis. In this case field configurations $H\simeq M$,
by assumption, do not cost prohibitive energy to excite.

In all cases, the effective theory faithfully reproduces the
vacuum behaviour of the underlying theory to which it is an
approximation. If the ground state energy in this underlying
theory has flat directions (or any generic field region) 
which cost little energy, then the
potential in the effective theory will be nonpolynomial, but
stable. If, on the other hand, the underlying energy  has no
shallow directions, the effective potential will be well
approximated by its expansion in powers of $H$ about the vacuum.

{\bf 3. A Revised Analysis:} Based on the insights suggested by
the toy model, we may now reanalyze the stability analysis within
the SM supplemented by NROs in the Higgs potential. For the
purposes of the analysis we adopt the point of view that, sometime
into the future, the Higgs particle has been discovered and the
value of the Higgs mass is known. We also assume that although no
new physics beyond the SM has been observed directly, Higgs
scattering is sufficiently well measured that there is indirect
evidence for new physics through a non-renormalizable $|H|^6$
interaction in the Higgs potential.

In this scenario the scale $M$, need not {\it a priori} be related
to the scale $\Lambda^*$ suggested by the SM stability analysis.
We therefore consider three different cases according to whether
$M$ is smaller than, comparable to, or larger than the instability
scale $\Lambda^*(M_h)$ in the pure SM. This is in contrast with
the analyses in refs.~\cite{datta,wudka}, which always assume
$M\simeq \Lambda$.

In all discussions about the relative sizes of the scales $M$ and
$\Lambda^*$ one must bear in mind that the low-energy theory does
not allow a separate determination of $M$ and the coupling
$\alpha$ which appear in the combination $\alpha \, |H|^6/M^2$.
For instance, if such a sextic term were generated at tree level
by the exchange of a heavy scalar, $S$, having a coupling $g S
H^3$, then $\alpha \sim g^2$. If, however, it is generated at one
loop by the circulation of a heavy fermion of mass $M$ we would
expect $\alpha \sim y^6/(4\pi)^2$, where $y$ is the
heavy-particle/Higgs Yukawa coupling. Strongly coupled physics
might generate $\alpha \sim {\cal O}(1)$. Clearly in
the cases where $\alpha \ll 1$, the scale indicated by the
coefficient of the $|H|^6$ term, $M/\sqrt{\alpha}$, can be
much larger than the the actual mass, $M$, of the virtual
particle which is responsible for this term.

{\bf 3.1. The case }${\bma M}{\bma \ll}{\bma \Lambda}^{\bma
*}{\bma (}{\bma M}_{\bma h}\bma{)}$. This is interesting by
itself, to the extent that the new physics that generates $|H|^6$
is much closer to the electroweak scale than anticipated by the
instability analysis. Of course, the fact that new physics appears
below $\Lambda^*$ is consistent with the expectations derived from
the pathological behaviour of the potential at $\Lambda^*$ ({\it
i.e.} some new physics at, or below, $\Lambda^*$ should cure that
pathology).

In any case, once there is evidence for new physics at scales $M
\ll \Lambda^*$ coupled to the Higgs sector, there need not be any
new physics at the scale $\Lambda^*$ at all. After all, our only
evidence that something happens at $\Lambda^*$ relies on using the
low-energy Higgs potential in a stability analysis at this scale,
and we have no justification for believing this potential is a
good approximation above the scale $M$. A reliable determination
of the instability scale must wait until the theory that describes
the physics at scales of order $M$ is worked out. We see that the
case $M\ll \Lambda^*$ eliminates the instability question from the
list of real problems of the model, in the same sense that the
appearance of a Landau pole in the electromagnetic coupling at a
scale much larger than the Planck scale is not a real problem in
the SM.

Once the new physics at $M$ is specified, then one can examine
whether it modifies the stability bounds or not. This was done,
for example, in Ref.~\cite{ff} for the case in which the new
physics is a heavy fourth family and in Ref.~\cite{nub} for the
case of heavy right-handed neutrinos that implement a see-saw
mechanism.

{\bf 3.2. The case }${\bma M}{\bma \simeq}{\bma \Lambda}^{\bma
*}{\bma (}{\bma M}_{\bma h}\bma{)}$. This case would be
interesting because one would have two independent indications for
new physics at the same scale. To the extent that the additional
Higgs NROs enter with coefficients which exacerbate the low-energy
stabilities -- such as a term $- \alpha \, H^6$ with $\alpha
> 0$ -- then we know that the vacuum energetics drives $H$ out to
values of order $M$, for which the expansion in powers of $H/M$
breaks down.

It is difficult to go much further than this, however, because the
breakdown of the $1/M$ expansions removes our only tool for
analysis. One needs to know the physics at $M$ to ascertain
whether there is an instability in the potential, and at what
scale it occurs. This is the low-energy theory's way of telling us
that we are asking a question -- the value of $H$ which minimizes
the energy -- whose answer cannot be reliably obtained purely
within an effective theory without reference to the degrees of
freedom at scale $M$.

{\bf 3.3.} ${\bma M}{\bma \gg}{\bma \Lambda}^{\bma *}{\bma (}{\bma
M}_{\bma h}\bma{)}$. This is the one case for which purely
low-energy calculations can reliably determine the influence, if
any, of the $|H|^6$ term on the calculation of $\Lambda^*$. We
must recognize that the limit $M \gg \Lambda^*$ is somewhat
strange, because we know physics at scales $\Lambda^*$ must
ultimately allow us to understand the stability of the
electroweak-breaking minimum of the Higgs potential, but we are
assuming this is {\it not} done by growing effective $|H|^6$ terms
with coefficients of order $(1/\Lambda^{*})^2$. If we have
separate indications for new Higgs-related physics at a larger
scale $M$, the odds are that such new physics is unrelated to the
physics that cures the instability problem.

Although $M \gg \Lambda^*$ is odd, we do not believe it is
absolutely impossible. For instance, one might imagine there to be
a selection rule in the low-energy theory which forbids the
generation of the $|H|^6$ term at the scale $\Lambda^*$, but not
at higher scales $M$ (in much the same way that flavour-changing
interactions amongst fermions arise in the low-energy limit in the
Standard Model). This might not preclude the appearance at scale
$\Lambda^*$ of higher powers of $|H|$ than $|H|^6$, which would be
the precursors in the low-energy potential of the stabilization
which happens at this scale. It is difficult to see how a
selection rule could do this for an $|H|^6$ term (for a single
Higgs field) taken in isolation, but it might conceivably arise
within a supersymmetric theory, where the scalar potential is
generated by a superpotential which can be subject to selection
rules and nonrenormalization theorems at scale $\Lambda^*$.

In any case, putting aside the issue as to how the low-energy
theory arises, we adopt the spirit used in the literature, and
simply examine how the prediction for $\Lambda^*$ is changed in
the low-energy theory by the appearance of an effective $|H|^6$
term.

To see how $\Lambda^*(M_h)$ gets modified by $\alpha\neq 0$ we
have to determine the balance between the $|H|^4$ and $|H|^6$
terms in the potential. For $|H|\sim\Lambda^*$, the quartic term
is \be \delta_4V={1\over 2}[\lambda(Q)+\delta\lambda]|H(Q)|^4\ ,
\ee where $Q$ must be chosen as $Q\sim|H|$ and the $Q$-dependence
of $\lambda$ and $H$ is governed at one-loop by\footnote{One can
add the effects of non-renormalizable operators to these
renormalization group equations (see \cite{wudka}). They represent
a small correction for our purposes and we ignore them here. }
\be
{d\lambda\over d \log Q}\equiv \beta_\lambda = {1\over
16\pi^2}\left[12(\lambda^2-h_t^2+\lambda
h_t^2)-(9g^2+3{g'}^2)\lambda
 +{3\over 4}(3g^4+2g^2{g'}^2+{g'}^4)\right]\ ,
\label{betal}
\ee
and
\be
{d\log |H|\over d \log Q}\equiv   \gamma_H
={1\over 16\pi^2}\left[{3\over 4}(3g^2+{g'}^2-4h_t^2)\right]\ ,
\ee
where $h_t$ is the top-quark Yukawa coupling, and $g,g'$ are the
$SU(2)_L$
and $U(1)_Y$ gauge couplings, respectively. We find that, for field values
in the neighbourhood of $\Lambda^*$ [defined by
$\lambda(\Lambda^*)+\delta\lambda=0$], the quartic term of the
potential is
given by
\be
\delta_4V={1\over 2}
\beta_\lambda(\Lambda^*)|H|^4\left[1+4\gamma_H(\Lambda^*)\log
{|H|\over\Lambda^*}\right]\log{|H|\over\Lambda^*}\ ,
\ee
while the non-renormalizable part will have the form
\be
\delta_6V=-{1\over 3M^2}|H|^6
\left[\alpha(\Lambda^*)+\beta_\alpha(\Lambda^*)\log{|H|\over\Lambda^*}\right]
\left[1+6\gamma_H(\Lambda^*)\log
{|H|\over\Lambda^*}\right]\ ,
\ee
where $\beta_\alpha\equiv d\alpha/d\log Q$ is given in ref.~\cite{wudka}.

%%C I wasn't sure what to put here, what did you have in mind?
%{\bf Comment on the possible differences between
%$\alpha(\Lambda^*)$ and $\alpha(M)$. But this probably belongs in
%the previous section. [Cliff?]}

The new instability scale $\Lambda^*_\alpha\equiv\Lambda^*(M_h,\alpha)$
is determined by the condition $\delta_4V+\delta_6V\simeq 0$, which leads
to
\be
\beta_\lambda(\Lambda^*)\log{\Lambda^*_\alpha\over\Lambda^*}
\simeq {2\over 3}\left({\Lambda^*_\alpha\over M}\right)^2
\left[\alpha(\Lambda^*)+(\beta_\alpha+2\alpha\gamma_H)
\log{\Lambda^*_\alpha\over\Lambda^*}
\right]\ ,
\ee
and this implies
\be
\Lambda^*_\alpha\simeq \Lambda^*\exp\left[
{2\over 3}{\alpha(\Lambda^*)\over\beta_\lambda}\left(
{\Lambda^*\over M}\right)^2
\right]\ .
\label{red}
\ee
Remembering that $\beta_\lambda\leq 0$ (this caused the instability in the
first place), eq.~(\ref{red}) shows that the instability scale is decreased
slightly by a non-zero positive $\alpha$ (of course, if $\alpha<0$, the
instability scale would increase a little).

The Higgs stability bound has a logarithmic sensitivity to $\Lambda$ and
therefore the reduction (\ref{red}) affects very mildly the value of this
bound. Its change can be estimated to be
\be
\delta {M_h^*}^2\simeq {\alpha(\Lambda^*)\over 3}\left(
{\Lambda^*\over M}\right)^2v^2\ ,
\label{dm}
\ee
where $v\simeq 246$ GeV is the Higgs VEV. For reasonable values of
$\alpha(\Lambda^*)$ this is a very small shift (much smaller than the
$\sim 5$ GeV uncertainty due to the indetermination of $\alpha_s$ or
$h_t$).

The
only possible caveat in the previous discussion is the possibility
of having $\beta_\lambda(\Lambda^*)\simeq 0$. This clearly requires very
particular values of the parameters, but in fact it can be arranged, as we
explain in what follows. As we have seen, for low enough  values of $M_h$,
$\beta_\lambda(M_W)$ is negative, due to the $-h_t^2$ term in
(\ref{betal}), and drives $\lambda$ towards negative
values at higher scales. However, $h_t$ decreases with increasing energy,
and eventually $\beta_\lambda$ turns positive again. This does not solve
the instability problem, though, because if $\lambda$ takes negative values
in some energy range the potential develops a very deep minimum there and
cannot be accepted, even if $\lambda$ is positive at even higher energies
\cite{stab}.

%%%%%%%%%%%%%%%%%%%%%%%%figure%%%%%%%%%%%%%%%%%%%%%%%%
\begin{figure}
%\psdraft
\centerline{
\psfig{figure=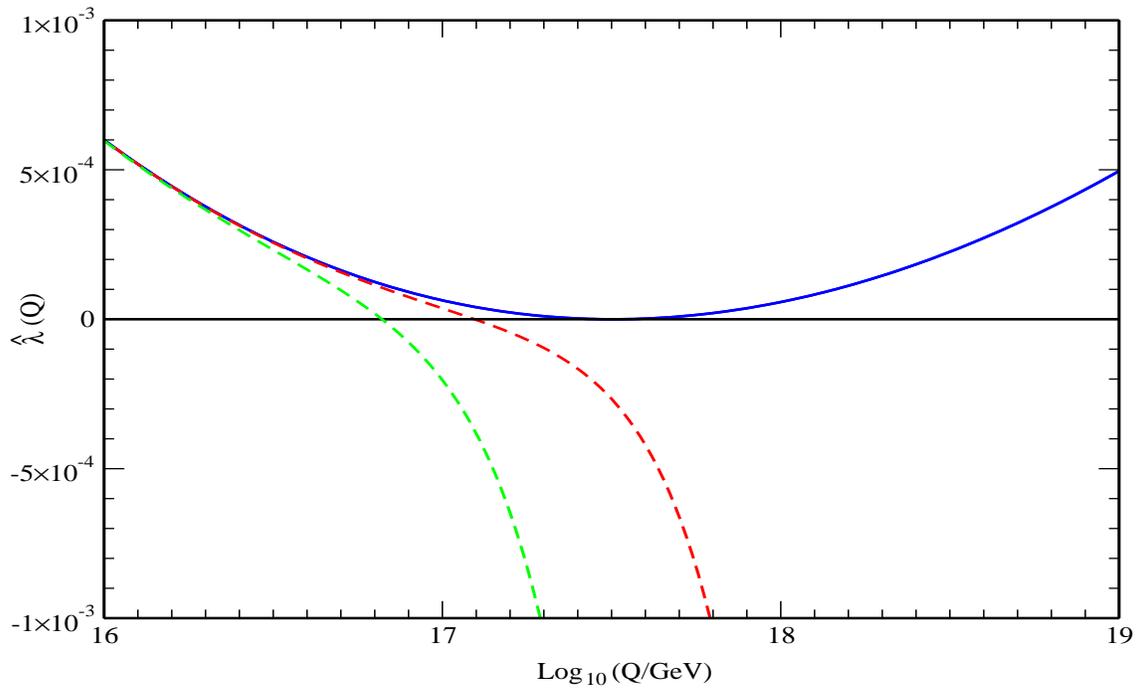,height=10cm,width=10cm,angle=-90,bbllx=1.cm,%
bblly=5.cm,bburx=20.cm,bbury=21.cm}}
\caption
{\footnotesize Running $\hat\lambda_\alpha(Q)$ for the pure SM
($\alpha=0$, solid line) and for the SM with a NRO like (\ref{NRO}) with
$M=M_p$ and $\alpha=0.1$, 1 (dashed lines). Other parameters are: $M_t=169.2$ GeV,
$\alpha_s(M_Z)=0.12$ and $M_h=121.5$ GeV.}
\end{figure}
%%%%%%%%%%%%%%%%%%%%%%%%%figure%%%%%%%%%%%%%%%%%%%%%%%%

%%%%%%%%%%%%%%%%%%%%%%%%figure%%%%%%%%%%%%%%%%%%%%%%%%
\begin{figure}
%\psdraft
\centerline{\vbox{
\psfig{figure=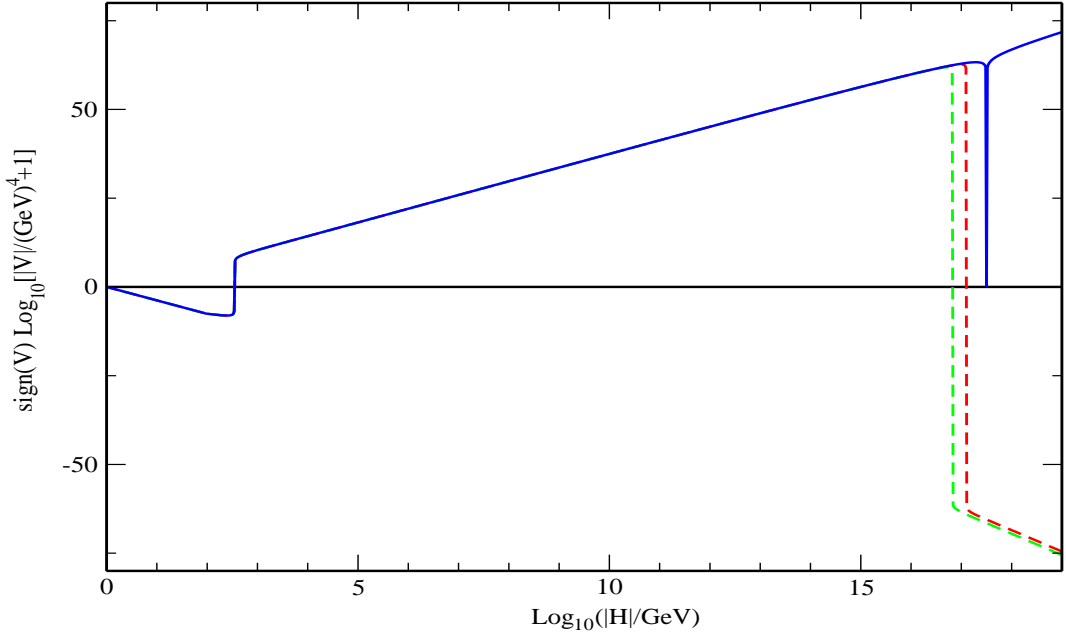,height=10cm,width=10cm,angle=-90,bbllx=3.cm,%
bblly=6.cm,bburx=23.cm,bbury=23.cm}
\psfig{figure=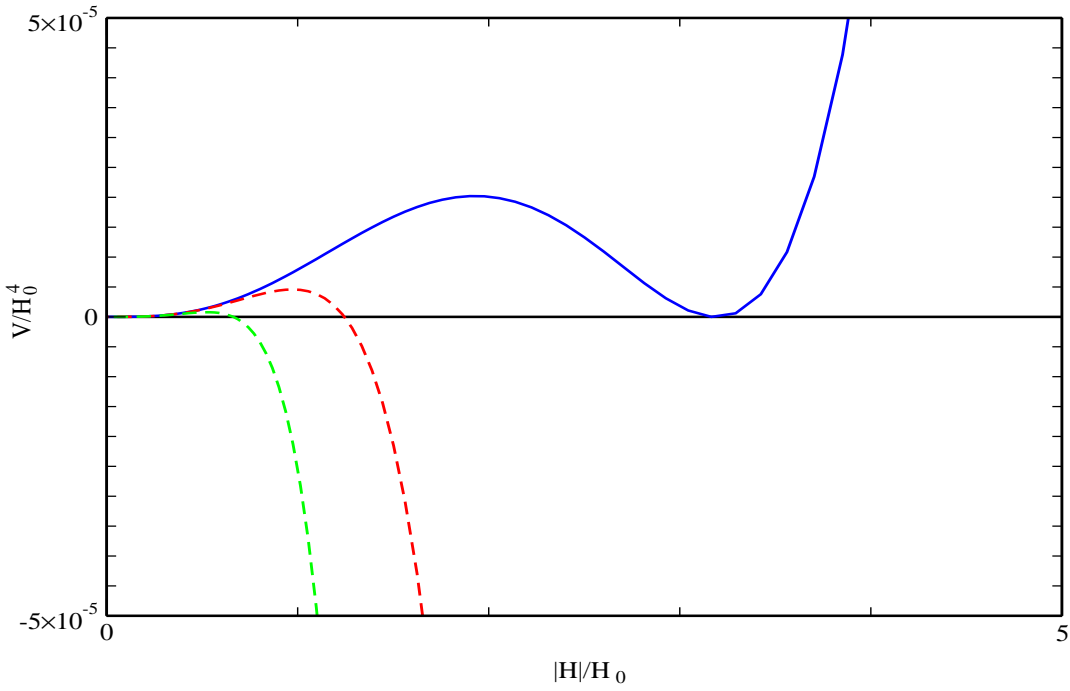,height=10cm,width=10cm,angle=-90,bbllx=1.cm,%
bblly=6.cm,bburx=20.cm,bbury=23.cm}}}
\caption
{\footnotesize Upper plot: Effective potential corresponding to the
parameters of fig.~1, with the same line coding. Lower plot: detail of
the instability region.}
\end{figure}
%%%%%%%%%%%%%%%%%%%%%%%%%figure%%%%%%%%%%%%%%%%%%%%%%%%

In any case, what matters now is that, in cases like the ones described in
the previous paragraph, there is an energy scale
at which $\beta_\lambda(Q_0)\simeq 0$. The case we are after is that in
which $\beta_\lambda(Q_0)\simeq 0$ {\it and} $\lambda(Q_0)\simeq
0$, {\it i.e.}
\be
\beta_\lambda(\Lambda^*)\simeq 0\ .
\label{condit}
\ee
In table~1, we list the values of $M_h$ and $\Lambda^*$ for which
(\ref{condit}) holds,
taking $M_t=174.3\pm 5.1$ GeV and $\alpha_s(M_Z)=0.118\pm 0.002$, the
current experimental ranges as given in ref.~\cite{PDG}.
For all our numerical work, we use the renormalization-group-improved
one-loop Higgs potential, with parameters running with two-loop
beta-functions (see Ref.~\cite{stab}). From table~1 we see that
condition (\ref{condit}) can only be arranged for rather large
values of $\Lambda^*$, not far from the Planck Mass. Entries
with $\Lambda=M_p$ in table~1 correspond to cases that do not satisfy
condition (\ref{condit})
for any scale below $M_p$. (In other words, for such parameters
one would get $\Lambda^*>M_p$.) The value of $M_h$ quoted for these cases
is simply the stability bound for $\Lambda=M_p$. Figure~1 shows the
running of
$\hat\lambda(Q)$ (solid line) for the extreme case $M_t=169.2$ GeV and
$\alpha_s=0.12$.
It is clear that, for values of $M_h$ higher than  $M_h^*\simeq 121.5$
GeV (see table~1),
the SM potential is free of instabilities at any scale (for these
values of $M_t$ and $\alpha_s$). In
this sense, the stability bound associated with cut-off scales $\Lambda$
beyond $\Lambda^*\sim 8\times 10^{16}$~GeV is independent of the value
of $\Lambda$.
This implies, in other words, that [for the previous values of $M_t$ and
$\alpha_s(M_Z)$]
 the potential of the SM with cut-off $\Lambda=10^{19}$~GeV,
and $M_h$ slightly below $M_h^*\simeq 121.5$ GeV develops an instability
around
$\Lambda^*\sim 8\times 10^{16}$~GeV, well below the cut-off
$\Lambda=10^{19}$~GeV.
\[
\]

\begin{center}
\begin{tabular}{||c|c|c|c||}\hline
$\alpha_s(M_Z)\downarrow$\, ; $M_t$[GeV]$\rightarrow$ & 169.2 & 174.3 &
179.4
\\ \hline
0.120 & 121.5; $3\times 10^{17}$ &  132; $M_p$ &
142.5; $M_p$ \\ \hline
0.118 & 123; $9\times 10^{17}$ &  133.5; $M_p$ &
144; $M_p$ \\ \hline
0.116 & 125; $3\times 10^{18}$  & 135; $M_p$  & 145; $M_p$  \\
\hline
\end{tabular}
\vspace{0.5cm}
\end{center}
\begin{center}
Table 1: {\footnotesize Values of $\{M_h; \Lambda^*\}$ in GeV, for the
indicated values of $\alpha_s(M_Z)$ and $M_t$.}
\end{center}

Assuming then that we live in such a world, with $M_t=169.2$~GeV,
$\alpha_s(M_Z)=0.12$ and $M_h\simeq 121.5$~GeV, the derivation
of $\Lambda^*_\alpha$ has to be more precise than the one 
presented before. In particular, we
have to keep higher order corrections in the evaluation of
$\hat\lambda(Q)$ which, around $\Lambda^*$, is given by 
\be
\lambda(Q)\simeq \beta_\lambda(\Lambda^*)\log {Q\over\Lambda^*}
+{1\over 2}\beta'_\lambda(\Lambda^*)\left[\log{Q\over\Lambda^*}
\right]^2 +... \label{lesp} 
\ee 
where $\beta'_\lambda\equiv
d\beta_\lambda/d \log Q$. The first term in the right hand side
is now zero, due to our choice of parameters [that ensure
$\beta_\lambda(\Lambda^*)=0$]. Using (\ref{lesp}), we can compute
the new instability scale for non-zero $\alpha$ and
find\footnote{We neglect here [as well as in (\ref{red})] the
scale dependence of $\alpha$.} \be \Lambda_\alpha^*\simeq
\Lambda^*\exp\left[-{\Lambda^*\over
M}\sqrt{{2\alpha(\Lambda^*)\over
3\beta'_\lambda(\Lambda^*)}}\right] \ , \ee while (\ref{dm}) is
still valid. This shows that the decrease in the instability scale
can be larger now, even if the change in the stability bound is
still very small.

Figure~1 shows also in dashed lines the running of
$\hat\lambda_\alpha(Q)\equiv \hat\lambda(Q)-2\alpha|H|^2/(3M^2)$,
quantity that governs the stability of the potential for non-zero
$\alpha$. We show two cases with $\alpha=0.1$ and $1.0$ (the curve
for $\alpha=0.1$ is closer to the solid, $\alpha=0$, line).
Figures~2a and 2b show the potential corresponding to the same
choices of parameters made in fig.~1. Fig.~2a presents the
potential in a form suitable to show its structure at all scales
({\it i.e.} the electroweak vacuum and the non-standard
minimum\footnote{We define $\Lambda^*$ by the condition $V=0$
in the non-standard minimum while we should have demanded degeneracy
of both minima. We use $V=0$ for simplicity given that the degeneracy
condition is more complicated to implement and the numerical
difference (say for $M_h^*$ and $\Lambda^*$) between both choices
is negligible.},
or instabilities, at $\Lambda^*\sim 10^{17}$ GeV). The line coding
is as in fig.~1. Figure~2b focuses on the structure of the
potential near the instability scale and presents simply $V/H_0^4$
versus $|H|/H_0$ with $H_0\equiv 10^{17}$ GeV. Both plots show clearly
the instabilities that appear for non-zero $\alpha$, but it is
manifest that, even in this particularly sensitive case, the shift
$\Lambda^*\rightarrow \Lambda_\alpha^*$ is not larger than an
order of magnitude. Alternatively, the stability bound associated
to the cut-off scale $\Lambda\simeq 3\times 10^{17}$ GeV for the
case $\alpha\neq 0$ (notice that, as explained in previous
sections, we cannot compute reliably the stability bound for
$\Lambda=M_p$ in this case) is nearly indistinguishable from the
$\alpha=0$ bound. We have checked that the difference is smaller
than 1 GeV.

In summary, we have re-examined the issue of how effective
non-renormalizable interactions in the Higgs potential, induced by
new physics, can change the understanding of vacuum stability
within the low-energy effective theory. We find that calculations
which remain within the effective theory's domain of validity
predict only small changes to the instability scale which is
inferred purely within the SM.

Because this conclusion disagrees with some previous analyses of
these same issues, we have re-examined these analyses and find
that they depend on the evaluation of $V$ for fields large enough
to invalidate its expansion in powers of $H$. This happens because
the large effect is claimed for non-renormalizable interactions
whose sign exacerbates any low-energy stability problems. As a
result these interactions tend to drive the potential minimum out
to fields which are too large to use truncated potentials.

\section*{Acknowledgements}

C.B.'s research is supported in part by funds from N.S.E.R.C.
(Canada) and F.C.A.R. (Qu\'ebec). V.D.C would like to thank PPARC for a 
Research Associateship.

%%%%%%%%%%%%%%%%%%%%%%%%%%%%%%%%%%%%%%%%%%%%%%%%%%%%%%%%%%%%%%%%%%%

\end{document}